\documentclass[aerospace,article,submit,pdftex,moreauthors]{Definitions/mdpi} 
\preto{\abstractkeywords}{\nolinenumbers}
\firstpage{1} 
\pubvolume{1}
\issuenum{1}
\articlenumber{0}
\pubyear{2023}
\copyrightyear{2023}
\datereceived{ } 
\daterevised{ } 
\dateaccepted{ } 
\datepublished{ } 
\hreflink{https://doi.org/} 

\usepackage[acronym]{glossaries}
\usepackage[caption=false,font=normalsize,labelfont=small,textfont=small]{subfig}
\usepackage{tabularx}
\usepackage{graphicx}
\usepackage{stfloats}
\usepackage{comment}
\usepackage{array}
\newcolumntype{x}[1]{>{\centering\arraybackslash\hspace{0pt}}p{#1}}
\makeglossaries

\newacronym{afdx}{AFDX}{Avionics Full-Duplex Switch Ethernet}
\newacronym{tteth}{TTEthernet}{Time-Triggered Ethernet}
\newacronym{tsn}{TSN}{Time Sensitive Networking}
\newacronym{ima}{IMA}{Integrated Modular Avionics}
\newacronym{vl}{VL}{Virtual Links}
\newacronym{bag}{BAG}{Band Allocation Gap}
\newacronym{oa}{OA}{Orthogonal Arrays}
\newacronym{snr}{SNR}{signal-to-noise ratio}
\newacronym{tm}{TM}{Taguchi Method}
\newacronym{anova}{ANOVA}{Analysis of Variance}
\newacronym{qos}{QoS}{Quality of Service}
\newacronym{dima}{DIMA}{Distributed IMA}

\newacronym{crc}{CRC}{Cyclic Redundancy Check}
\newacronym{fifo}{FIFO}{First In First Out}

\newacronym{udp}{UDP}{User Datagram Protocol}
\newacronym{ip}{IP}{Internet Protocol}

\newacronym{doe}{DoE}{Design of Experiments}
\newacronym{vv}{V\&V}{Validation \& Verification}

\newacronym{tc}{TC}{Termination Criteria}

\newacronym{es}{ES}{End Systems}

\newacronym{stb}{STB}{the-smaller-the-better}
\newacronym{ntb}{NTB}{the-nominal-the-best}
\newacronym{ltb}{LTB}{the-larger-the-better}

\newacronym{fom}{FoM}{Figures of Merit}

\newacronym{ber}{BER}{Bit Error Rate}

\newacronym{be}{BE}{Best Effort}
\newacronym{sct}{SCT}{Safety-Critical Traffic}
\newacronym{bls}{BLS}{Burst Limiting Shaper}

\newacronym{nsga}{NSGA-II}{Non-dominated Sorting Genetic Algorithm}

\newacronym{ga}{GA}{Genetic Algorithm}
\newacronym{ep}{EP}{evolutionary programming}

\newacronym{epl}{EPL}{Exact Possible Longest}
\newacronym{bncog}{BNCOG}{Burst kept Network Calculus with Optimized Grouping Strategy}

\newacronym{captor}{CAPTOR}{advanCed Avionics communications validation and verification PlaTfORm}

\newacronym{ai}{AI}{Artificial Intelligence}

\newacronym{mbse}{MBSE}{Model Based Systems Engineering}
\newacronym{sil}{SIL}{Software-in-the-loop}
\newacronym{hil}{HIL}{Hardware-in-the-loop}

\newacronym{ats}{ATS}{Asynchronous Traffic Shaper}
\newacronym{tas}{TAS}{Time Aware Shaper}
\newacronym{frer}{FRER}{Frame Replication and Elimination for Reliability}
\newacronym{cu}{CU}{Calculator Unit}

\newacronym{lan}{LANs}{Local Area Networks}
\newacronym{csmacd}{CSMA/CD}{Carrier Sense Multiple Access with Collision Detection}
\newacronym{mac}{MAC}{Media Access Control}
\newacronym{cots}{COTS}{Commercial off-the-shelf}

\Title{An event-driven link-level simulator for validation of AFDX and Ethernet avionics networks}

\TitleCitation{An event-driven link-level simulator for validation of AFDX and Ethernet avionics networks}

\Author{Pablo Vera-Soto $^{1}$\orcidA{}, Javier Villegas $^{1}$\orcidB{}, Sergio Fortes $^{1,}$*\orcidB{}, José Pulido $^{1}$\orcidB{}, Vicente Escaño $^{2}$\orcidB{}, Rafael Ortiz $^{2}$\orcidB{} and Raquel Barco $^{1}$\orcidB{}}

\AuthorNames{Pablo Vera-Soto, Javier Villegas, Sergio Fortes, José Pulido, Vicente Escaño, Rafael Ortiz and Raquel Barco}

\AuthorCitation{Vera-Soto, P.; Villegas, J.; Fortes, S.; Pulido, J.; Escaño, V.; Ortiz, R.; Barco, R.}

\address{%
$^{1}$ \quad Telecommunication Research Institute (TELMA), University of Málaga, E.T.S. Ingeniería de Telecomunicación, Boulevar Louis Pasteur 35, 29010, Málaga, Spain; \{pvera, jvc, sfr, josepa, rmb\}@ic.uma.es\\
$^{2}$ \quad Aerospace and Defence Systems, Aertec Solutions, 29590 Málaga, Spain; \{vescano, rortiz\}@aertecsolutions.com}

\corres{Correspondence: sfr@ic.uma.es}

\abstract{Aircraft are composed of many electronic systems: sensors, displays, navigation equipment and communication elements. These elements require a reliable interconnection, which is a major challenge for communication networks as high reliability and predictability requirements must be verified for safe operation. In addition, their verification via hardware deployments is limited because these are costly and make difficult to try different architectures and configurations, thus delaying the design and development in this area. Therefore, verification at early stages in the design process is of great importance and must be supported by simulation. In this context, this work presents an event-driven link level framework and simulator for the validation of avionics networks. The presented tool supports communication protocols such as Avionics Full-Duplex Switched Ethernet (AFDX), which is a common protocol in avionics, as well as Ethernet, used with static routing. Alsa, accurate results are facilitated by the simulator through the utilization of realistic models for the different devices. The proposed platform is evaluated in Clean Sky's Disruptive Cockpit for Large Passenger Aircraft architecture scenario showing capabilities of the simulator. The speed of the verification is a key factor in its application, so the computational cost is analysed, proving that the execution time is linearly dependent on the number of messages sent.}

\keyword{Avionics, ARINC664, AFDX, Ethernet, communications, verification and validation, protocols} 

\begin{document}
\section{Introduction}
\label{chp:intro}

The aerospace industry has undergone tremendous advances since its conception just over 100 years ago by the Wright brothers. Of great importance in these advances has been the introduction of \textit{avionics} (a term derived from the combination of \textit{aviation} and \textit{electronics}), which encompasses all the electronic systems that have been added to aircraft, including a wide range of equipment such as actuators, sensors and communication systems, and which make up the majority of the safety-critical elements in an aircraft. Within the advancement of avionics, the approach based on the concept of \gls{ima} \cite{Butz2008OpenIM} has been extended to commercial aviation with the design of airplanes like the Airbus A380 \cite{AirbusA380} and Boeing 777 \cite{Boeing777}. This approach involves distributing various safety-critical functions, which are becoming increasingly numerous due to technological advancements, into separate independent modules connected within an avionics network. Moreover, the new generations of \gls{ima}s use not only software distribution, as in the original versions, but also employ hardware distribution, which places the modules closer to the components they monitor for faster response. These newer generations are referred to as \gls{dima}. These systems cover diverse categories and applications, such as navigation, communication, flight control, etc.

There are several buses and protocols available for establishing this kind of networks, including Ethernet, CAN bus, and serial bus, among others. Ethernet-based protocols are currently the most widely used, with particular emphasis on \gls{afdx} \cite{Kazi2013ArchitectingOA}. \gls{afdx} is an implementation of the ARINC 664 Part 7 standard and offers dedicated bandwidth and a fixed \gls{qos}. Additionally, \gls{tsn} \cite{10.1145/3487330} standard is other Ethernet-based option that is expected to become the standard for future generations of aircraft. Currently, a working group is in the process of developing a \gls{tsn} profile tailored specifically to the avionics sector, covering aspects such as shapers, scheduling, and stream isolation \cite{TSN_TG}. Other protocols, such as \gls{tteth}, have also been proposed and are commonly used in spacecraft applications \cite{TTEthernet}.


Some protocols, such as \gls{afdx}, introduce a paradigm where determinism is achieved through a specific network structure. Consequently, each network configuration must undergo individual analysis to ensure that real-time requirements, including \gls{fom} like delay and jitter, are satisfied. Additionally, as avionics systems evolve, their bandwidth demands increase, making it challenging to devise new networks that align with safety and certification requirements. This validation process extends the time it takes to bring an aircraft to market. In this context, simulators play a crucial role in accelerating development.



\begin{figure}[H]
    \centering
    \includegraphics[width=0.7\textwidth]{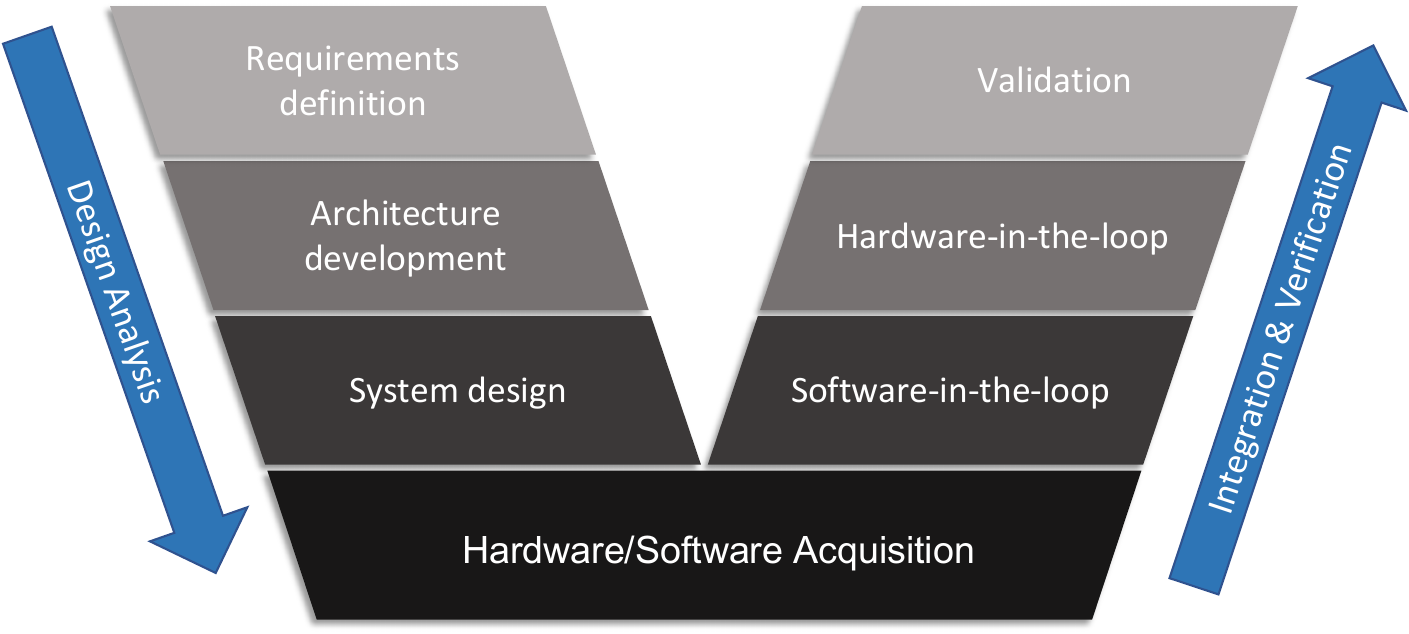}
    \caption{Systems engineering V-model.}
    \label{fig:mbse}
\end{figure}

In this context, \gls{mbse} \cite{wymore2018model} methodology is usually used in the design process of the aerospace sector \cite{Li2019}. This methodology uses models to design and analyse complex systems and involves the creation of a digital representation of a system. In this way, the model can be used to simulate and test different scenarios before the system is built. The \gls{mbse} design process consists in a realization of a series of steps, which are depicted in the V-model of Figure \ref{fig:mbse}. These steps range from requirements definition to \gls{sil}, \gls{hil} and integration. 

In this matter, the \gls{sil} phase is of great importance in the aerospace sector for several reasons. First and foremost, the significant cost associated with avionics equipment requires a cautious approach, making it impractical to move to a hardware implementation until certain design guarantees have been met. In addition, the complexity, heterogeneity, and vast number of elements within a network limit the amount of hardware testing that can be done prior to final implementation. Furthermore, stringent delay and throughput requirements impose the need to evaluate networks across different architectures and potential scenarios. Finally, the presence of a wide variety of protocols and technologies, including emerging ones such as \gls{tsn}, emphasises the need for a comprehensive evaluation during the \gls{sil} phase.

Considering this, the development of new approaches to support and speed up the evaluation of this kind of networks has been considered essential. In this field, a review of the main communication protocols used in avionics networks, has been carried out and an event-driven link level simulator has been designed and proposed for the validation of these networks, supporting the main protocols of Ethernet and \gls{afdx}. For this purpose, the elements of the network in the simulator have been designed, as well as the scenarios for the simulator evaluation. Finally, a validation of this simulator has been carried out.

Therefore, the present work is structured as follows. Section \ref{chp:Protocols} gives a brief summary of the supported protocols. Section \ref{chp:simulator} introduces the simulator developed. Section \ref{chp:Evaluation} analyses the correctness of the simulator results and the computational performance. Finally, Section \ref{chp:conclusions} presents the main conclusions of this work.

\section{Avionic protocols}
\label{chp:Protocols}

In this section, some of the most important protocols used in avionics networks nowadays are analyzed.

\subsection{ARINC 664}

\begin{figure}[h!]
    \centering
    \includegraphics[width=0.6\textwidth]{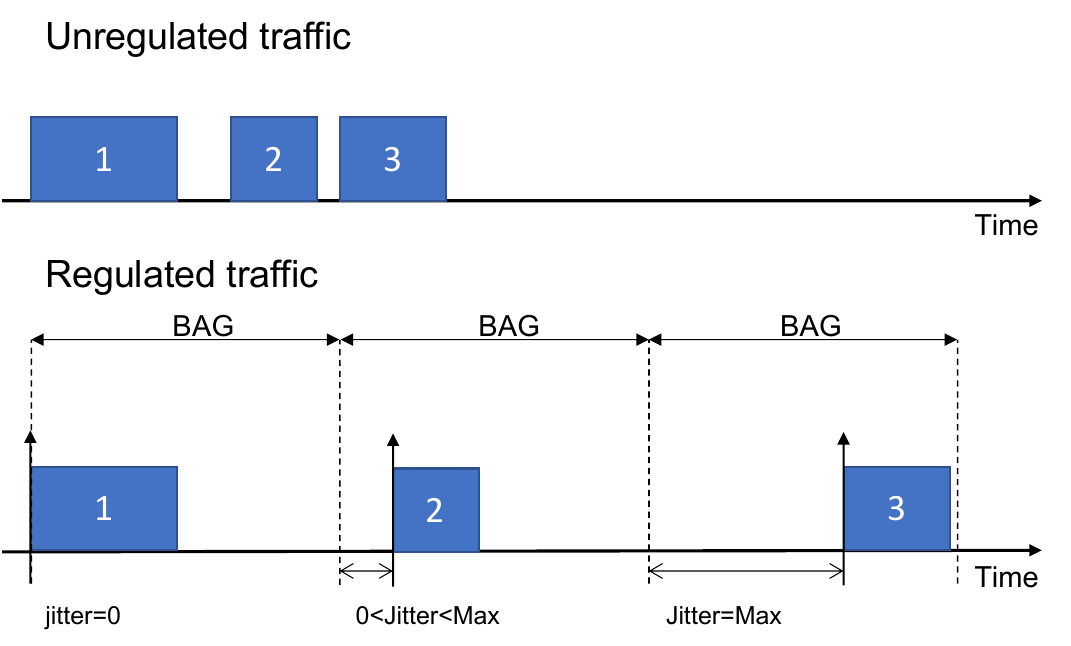}
    \caption{Traffic regulation in \gls{afdx} through \gls{bag}.}
    \label{fig:bag}
\end{figure}

\gls{afdx} is a packet switching protocol built over Ethernet networks that provide deterministic timing and redundancy management while using \gls{ip} and \gls{udp} as upper layer protocols. This protocol is characterised by three factors. First, the ARINC 664 standard proposes a duplicity in the network hardware for redundancy purposes. Second, \gls{afdx} implements the packet routing by defining unidirectional logical paths to follow called \gls{vl}. Third, \gls{afdx} regulates the packet traffic by limiting the bandwidth through the so-called \gls{bag}, which is the maximum rate at which data can be sent, and the data is guaranteed to be sent at that interval. Therefore, with the \gls{bag}, the maximum jitter is bounded, as depicted in Figure \ref{fig:bag}. These three factors are the ones that provides the determinism to the network.

\gls{afdx} networks are composed by two types of devices: \gls{es}, which are the end points of the network, and switches for interconnecting the \gls{es}.

\subsubsection{End System}

The \gls{es}s are entities that act as interface between the \gls{ima} equipment and the \gls{afdx} network, allowing the communication between them and multiplexing the data flows through the configured \gls{vl}s. ARINC 664 specifies network redundancy by implementing two identical networks to provide security in the communications. Therefore, the \gls{es}s send and receive the packets through two \gls{vl}s simultaneously. Therefore, the \gls{es} must manage the redundancy at the reception with a redundancy management policy, for which \textit{first-valid-wins} is usually used.

\subsubsection{Switch}

The switches are the entities responsible for delivering the packets to their destinations. In ARINC 664, the routing is done at the Data-Link layer, so the \gls{afdx} switches must be able to perform layer 2 switching, for which the \gls{vl} identifier in the frame header shall be used. Besides, the switches are responsible for applying the different policies in order to ensure determinism in the network.

\subsection{Ethernet for avionics}

\gls{afdx} is a widely spread protocol in aircraft. The redundancy that this protocol implements provides great reliability, but at the same time it supposes an increase in the number of devices and cables on board the aircraft, thus increasing the weight and integration costs of the aircraft. For instance, the Airbus A380 contains 500 km of cables \cite{Furse2002}. So, despite this protocol being commonly used in avionics, there are efforts to change the protocol implemented in order to reduce costs. In this way, Ethernet networks with static routing are still used in aircraft.

Ethernet is a widely used networking technology that forms the backbone of \gls{lan}. Developed in the 1970s, Ethernet has become a standard for wired networking, providing a reliable and efficient way for devices to communicate within a local area network. The basic principle of Ethernet is the use of a protocol that governs how data is transmitted over the network. It relies on a system of frames, which are packets of data that contain both source and destination addresses, allowing devices on the network to identify and process the information.

Ethernet typically uses a star or bus topology where devices are connected to a central hub or switch. Each device on the network has a unique identifier, known as a \gls{mac} address, which assists in the proper routing of data. The most common form of Ethernet uses twisted-pair cable with RJ45 connectors, and data is transmitted using a \gls{csmacd} protocol. With advancements such as Gigabit Ethernet and 10 Gigabit Ethernet, Ethernet continues to adapt to the increasing demands of modern networking, providing a robust and scalable solution for connecting devices within a local area network.

For example, \cite{Champeaux2016} proposes to get rid of the switches for small avionics network and build a distributed avionics communication network using Ethernet-based devices. Then, in \cite{Mifdaoui2018} and \cite{Amari2018}, new topologies are explored for Ethernet-based avionics networks, making emphasis in ring topologies. They compare the Airbus A380 \gls{afdx} topology with variations in an Ethernet ring topology, resulting in better delays than \gls{afdx} implementation. 

This effort to leaving \gls{afdx} behind can be seen in the market. For example, in \cite{Doverfelt2020}, an analysis of an \gls{afdx} and a custom Ethernet-based network is made in order to determine which implementation is more suitable for what a specific enterprise is looking for. This work determine that the Ethernet implementation is more flexible and suitable for the enterprise interests. However, Ethernet is not as reliable as \gls{afdx}, so a thoughtful validation must be done to ensure that they fulfill the real-time requirements needed, thus justifying simulators such as the one presented in this work.

\subsection{Time-Sensitive Networking}

As stated in \cite{Annighoefer2019}, it is expected that \gls{afdx}, whose \gls{cots} products have a high cost, will transition to the emerging \gls{tsn}, which can be implemented with low-cost \gls{cots} components, and it may be superior to \gls{afdx} in terms of determinism and performance.

\gls{tsn} is a set of standards of IEEE 802.1 \cite{IEEE802.1} based on Ethernet to provide communications with real time requirements. It includes several profiles, including Audio Video Bridging (802.1 BA), Fronthaul (802.1 CM/de), Industrial Automation (IEC/IEEE 60802) and Automotive In-Vehicle (P802.1 DG). Recently, \gls{tsn} has emerged as a promising protocol for avionics networks. The IEEE 802.1 Task Group is actively developing a \gls{tsn} profile specifically tailored for avionics networks (IEEE 802.1 DP), which need specifications slightly different from the other profiles.

To create the aerospace profile, the Task Group is replicating \gls{afdx} protocol, adapting its structure with IEEE 802.1 substandards. In contrast to the \gls{afdx} \gls{vl}, the IEEE 802.1Q Stream is introduced; the \gls{afdx} \gls{es} is opposed by the IEEE 802.1Q End Station, and the \gls{afdx} Switch is replaced by the IEEE 802.1Q Bridge.

On the one hand, various shapers are being considered, including \gls{ats} and synchronous alternatives such as credit-based shapers, \gls{tas} or \gls{bls}. Asynchronous shapers are designed for slower time cycles (periodicities greater than 50 ms), while synchronous shapers are designed for periodicities around 1 ms.

On the other hand, aerospace \gls{tsn} networks require redundancy similar to \gls{afdx}, which shall be implemented through the 802.1 CB - \gls{frer} substandard. These \gls{tsn} networks accommodate not only \gls{tsn} traffic, but also \gls{be} traffic, increasing network flexibility. In addition, \gls{afdx} filtering and policing concepts are incorporated into the \gls{tsn} aerospace profile.

A notable difference between \gls{tsn} and \gls{afdx} is the ease of configuration, with \gls{tsn} networks benefiting from simplified configuration using the YANG data models developed by IEEE.

\section{Proposed system} 
\label{chp:simulator}


As emphasized in Section \ref{chp:intro}, the acquisition of essential metrics to evaluate the performance of avionics networks is a fundamental task during their development, validation, and verification processes. The primary concern is to verify that the specified delay thresholds, which are critical to the proper operation of aircraft, are consistently met. An avionics network simulator has been developed to achieve this objective. A previous version of this simulator is presented in \cite{AfdxJavi}, which only supports \gls{afdx} protocol and has a less advanced switch model based only in \gls{fifo}s and without switch capacity monitoring. Also, the current version is easier to configure and to create use cases.

\subsection{General framework}

On the one hand, the simulator takes a series of inputs in order to create the model for simulation. These inputs, which are summarized in Table \ref{tab:Inputs}, include the simulation time, \gls{ber} and the topology of the network. The topology encompasses the protocol used (Ethernet or \gls{afdx}), the connections between the different elements in the network (through adjacency matrix), the routing of each flow/\gls{vl} (can be set manually or randomly), the periodicity/\gls{bag}, the length of the frames and some parameters of the switches such as switching delay and the internal memory.

\begin{table}[h]
\caption{Input configuration parameters.}
\label{tab:Inputs}
\centering
\begin{tabular}{lll}

\textbf{Parameter}&&\textbf{Fields}\\
\hline

Simulation time       && Duration in seconds\\
\hline
BER      && Bit Error Rate\\
\hline
Topology&&Protocol\\
&& Identifier\\
&& ESs\\
&& Route A\\
&& Route B\\
&& Cable length (m)\\
&& Link speed (bps)\\
&& BAG/Periodicity (ms)\\
&& Min/Max Packet Length (B)\\
&& Switch characteristics (delay and memory)\\

\end{tabular}
\end{table}

On the other hand, the simulator provides the following \gls{fom} as outputs, which are useful for validating the networks and for integrating the simulator into \gls{vv} frameworks:

\begin{itemize}
    \item \textbf{Delay}: includes the maximum, minimum, mean and standard deviation values of each flow/\gls{vl} in milliseconds. The delay is set as the time from departure to arrival.
    \item \textbf{Jitter}: includes the maximum, minimum, mean, standard deviation values of each flow/\gls{vl} in milliseconds.
    \item \textbf{Throughput}: includes the maximum, minimum, mean, standard deviation values of each flow/\gls{vl} in bits per second (bps).
    \item \textbf{Packet Loss}: of each flow/\gls{vl} in percentage.
    \item \textbf{Switch Capacity}: general capacity of each switch through the simulation in percentage.
\end{itemize}


Using these inputs, the network model generation process uses the information to build the network model that includes all specified \gls{es} and switches. In addition, the model links each \gls{vl} to its corresponding \gls{es} and establishes all essential connections between \gls{es} and switches.

The simulator has been developed in Matlab/Simulink, which performs event-driven simulations modelling the packets as entities. 
It has been developed this way so there is a step in the simulation compiler only when it is needed, so the simulations can be faster. 
Besides, the simulator works by modelling the Data Link Layer packet transmission by managing the packet entities in the \gls{es} and switches generated models. Therefore, a key factor in the simulator is the models used for these elements of the network, which are described next.


\subsection{End System model}

The \gls{es} is modelled with two components: a receiver and a transmitter. In the context of emulating on-board equipment, the transmitter generates the frames by incorporating a packet generator source, a redundancy management module, and a route selection module. This logical process is illustrated in Figure \ref{fig:Switch-model}. The Timer and Packet Generator modules substitute the \gls{ima} device in the network and generate the frames in its place. So, the operation starts with a timer that is set when the packets are generated. As the data passes through the system, the redundancy management module intervenes where necessary, duplicating messages to improve data integrity. These processed messages, together with any duplications, are then routed through the appropriate physical interfaces to reach their intended destinations. In addition, the \gls{es} sets the \gls{crc} based on the \gls{ber} input, modelling transmission errors.







\begin{figure}[h!]
    \centering
    \includegraphics[width = 0.9\textwidth]{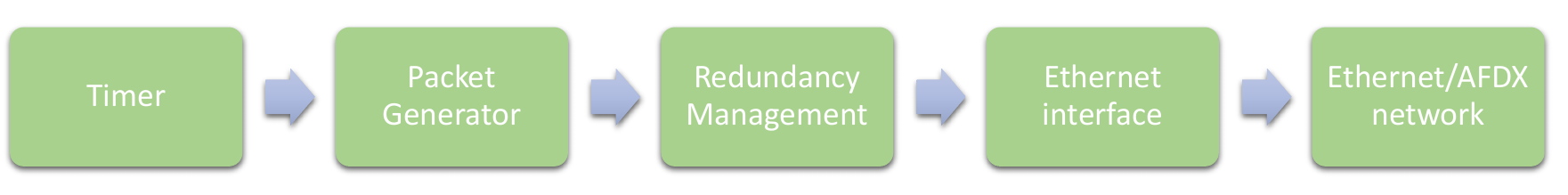}
    \caption{ES model in the simulator.}
    \label{fig:ES-model}
\end{figure}

\subsection{Switch model}

\begin{figure}[h!]
    \centering
    \includegraphics[width = 0.8\textwidth]{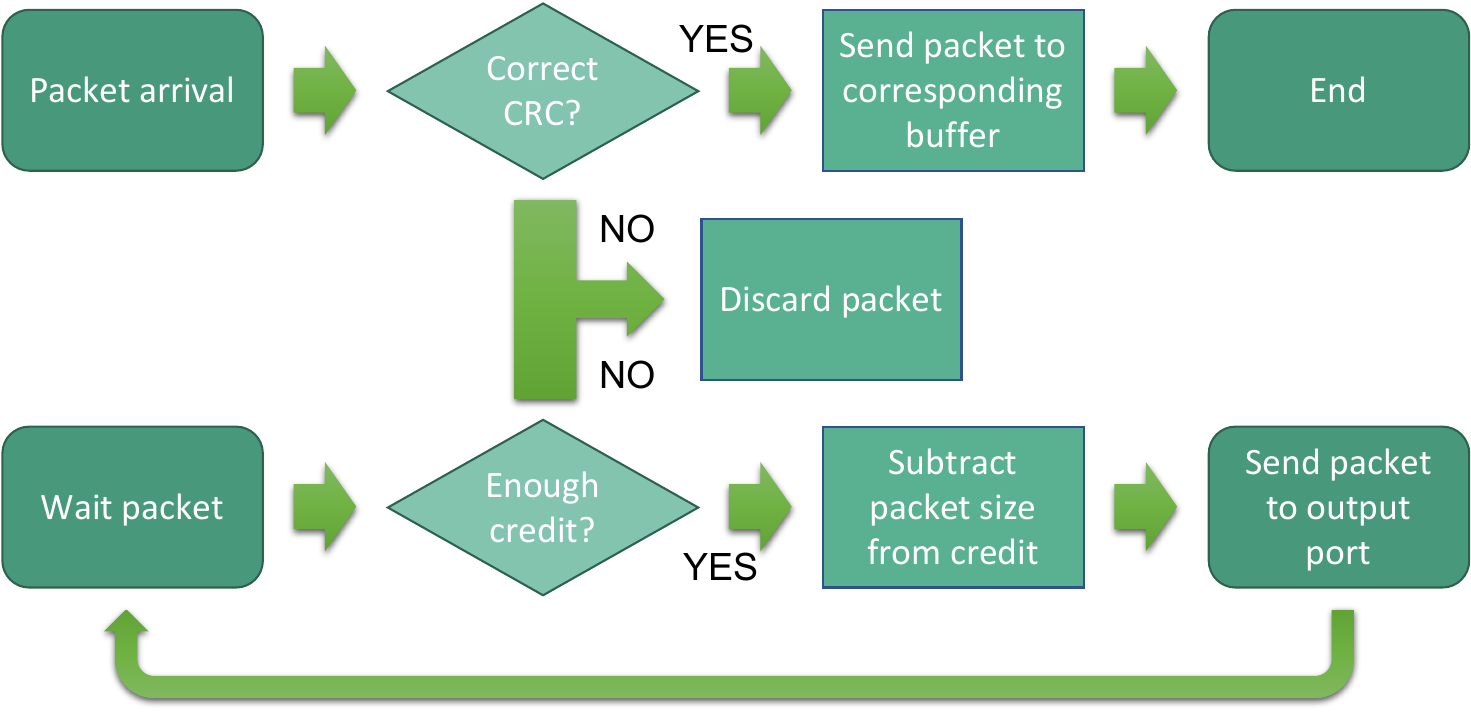}
    \caption{Switch logical model in the simulator.}
    \label{fig:Switch-model}
\end{figure}

The operation of the switch is structured around two core processes, namely scheduling and filtering policy, as shown in Figure \ref{fig:Switch-model}. The upper part of the figure is dedicated to scheduling, which uses the Round Robin algorithm to manage message prioritisation and transmission order. Conversely, the lower part of the figure focuses on the filtering policy, which is implemented using a Token Bucket shaper that regulates the rate of message flow. Within this switch, messages are discarded in case of incorrect \gls{crc} or insufficient credit.





\begin{figure}[h!]
    \centering
    \includegraphics[width = 0.6\textwidth]{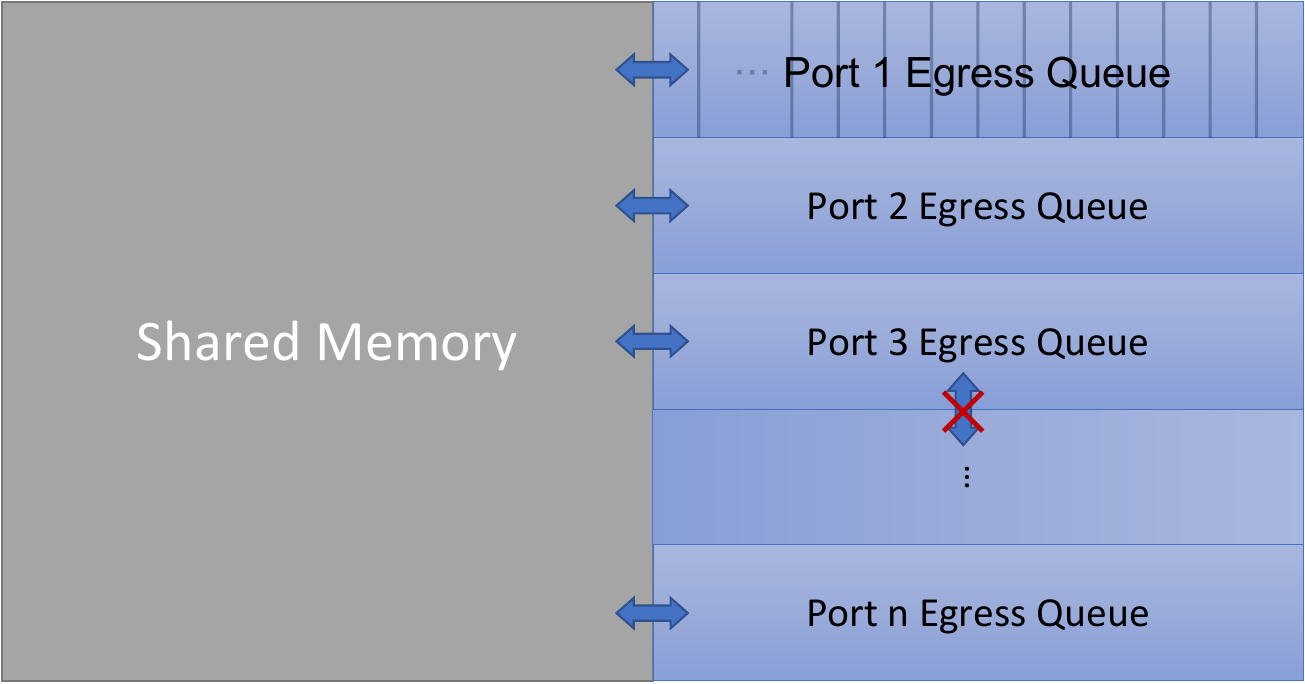}
    \caption{Architecture of the switch's memory.}
    \label{fig:Memory_arch}
\end{figure}

The internal memory configuration uses a shared queue system, a common approach found in commercial switches such as the one described in \cite{Switch}. As shown in Figure \ref{fig:Memory_arch}, this system consists of individual \gls{fifo} queues allocated to each port, providing dedicated memory space. In addition, there is a shared memory that is used when a particular \gls{fifo} reaches its capacity limit, ensuring that it does not compromise the reserved memory of other ports. This design protects each port from the saturation effects of burst traffic from other ports.


\subsection{Differences between protocol implementations}

Although \gls{afdx} is an Ethernet-based protocol, there are significant differences between them, as mentioned in Section \ref{chp:Protocols}. Therefore, there are notable differences that affect the handling of redundancy and routing strategies in the simulator implementation. While \gls{afdx} typically follows a standard approach that requires exact duplication of hardware components to ensure redundancy, Ethernet implementations offer a more flexible alternative. In terms of routing, Ethernet allows the use of IP routing and replaces the \gls{bag} with an assumed periodicity of messages that do not have to be a power of 2. This flexibility in Ethernet-based networks gives designers greater freedom in designing redundancy strategies. Consequently, the decision to use the redundancy management module within an \gls{es} becomes a matter of choice when working with Ethernet protocol in the simulator, as it may or may not be needed depending on the specific network design and redundancy requirements.





\section{Evaluation}
\label{chp:Evaluation}

Regarding the evaluation of the simulator, two main areas have been analysed in this work. First, the accuracy of the simulation results has been verified to ensure that the simulator could be trusted. Second, the computational performance has been analysed in order to determine the usefulness of the simulator, while at the same time providing an example of the results that the simulator can produce.

\subsection{Correctness of the results}

\begin{figure}[h!]
    \centering
    \includegraphics[width = 0.7\textwidth]{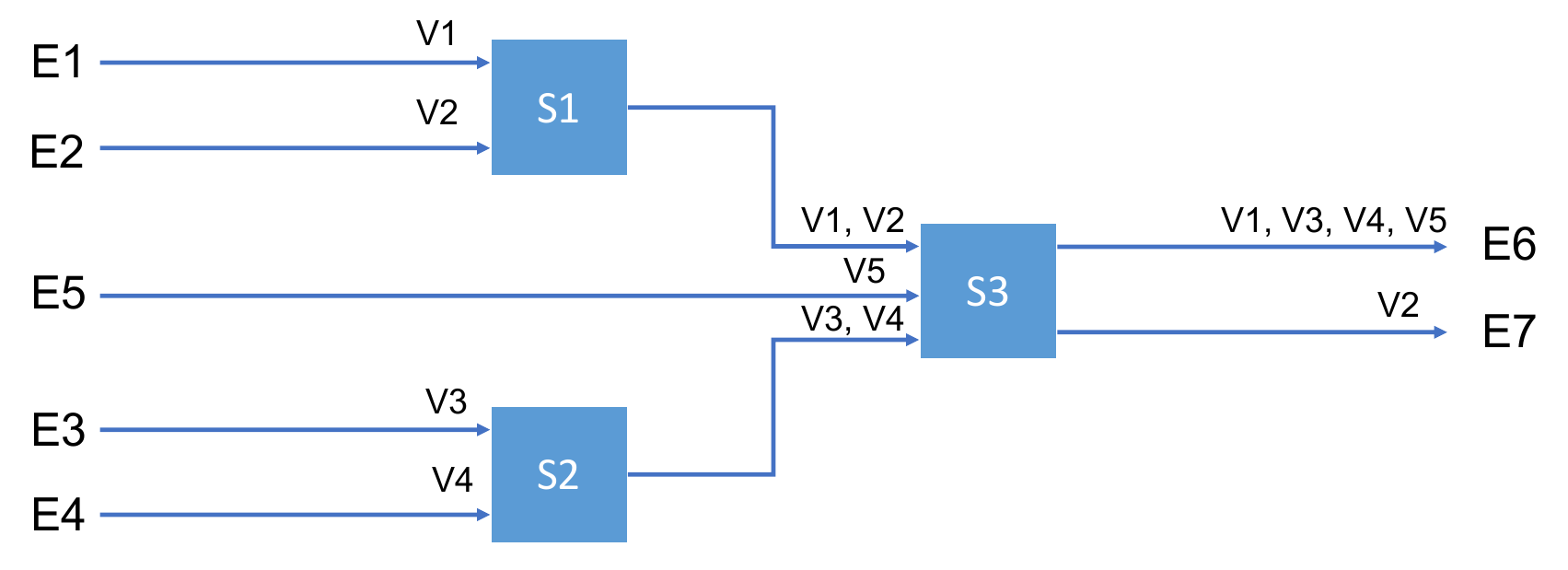}
    \caption{Testing use case topology adapted from \cite{Xu2019}.}
    \label{fig:Correct-arch}
\end{figure}

In order to check the correctness of the simulator results, a comparison with the work made in \cite{Xu2019} has been made. This work presents the use of an analytical method derived from Network Calculus for getting the worst possible delays in a \gls{afdx} network, then checks this method with a simple use case. The results of this use case have been replicated with the simulator presented in this work, in order to ensure that the simulator is capable of providing reliable results. The topology of the use case used is depicted in Figure \ref{fig:Correct-arch}.

\begin{figure}[h]
    \centering
    \includegraphics[width=\textwidth]{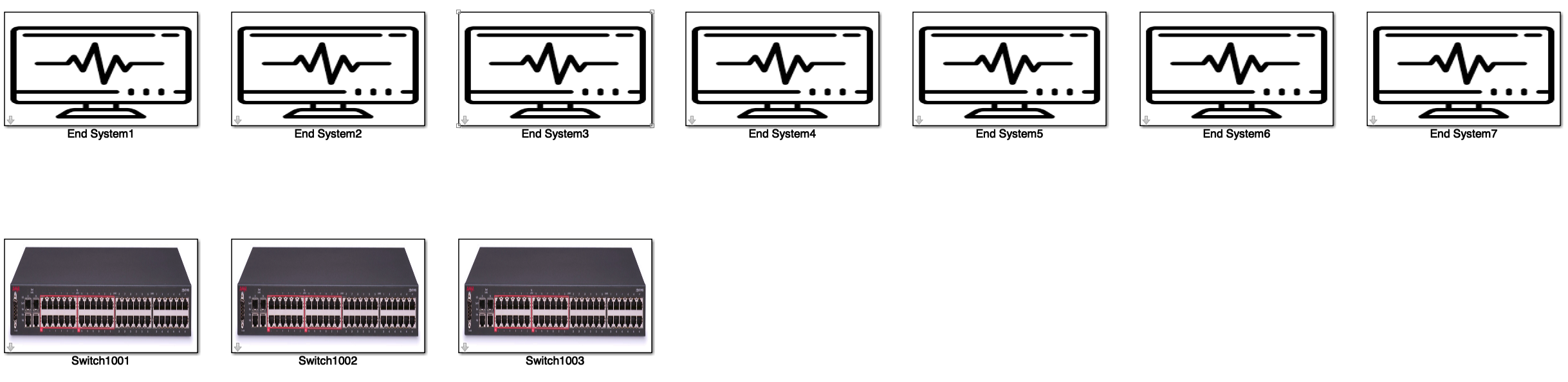}
\caption{Simulink model generated automatically.}
\label{fig:simulink}
\end{figure}

The topology is composed by 7 \gls{es}, 3 switches and 5 \gls{vl}s. The configuration of the \gls{vl}s in this use case is described in Table \ref{tab:check_paths}, where it can be observed that 2 \gls{vl}s go through switch S1, another 2 \gls{vl}s go through switch S2 and all 5 \gls{vl}s go through switch S3. Also, the length of the packets sent is 500 Bytes and the BAG configured is 4 ms. It can be seen in Figure \ref{fig:simulink} the Simulink model that the simulator has automatically generated for this use case.

\begin{table}[t]
\centering
\caption{Testing use case configuration \cite{Xu2019}.}
\label{tab:check_paths}
\resizebox{0.8\columnwidth}{!}{%
\begin{tabular}{|c|c|c|c|c|c|}
\hline
\rowcolor[HTML]{9B9B9B}
{\color[HTML]{FFFFFF}\textbf{Transmitter}} & {\color[HTML]{FFFFFF}\textbf{VL} }& {\color[HTML]{FFFFFF}\textbf{Receiver}} & {\color[HTML]{FFFFFF}\textbf{Path}} & {\color[HTML]{FFFFFF}\textbf{\begin{tabular}[c]{@{}c@{}}Packet\\ Length\end{tabular}}} & {\color[HTML]{FFFFFF}\textbf{BAG}} \\ \hline
ES1 & V1 & ES6 & ES1 $\xrightarrow{}$ S1 $\xrightarrow{}$ S3 $\xrightarrow{}$ ES6 & 500 B & 4 ms \\ 
ES2 & V2 & ES7 & ES2 $\xrightarrow{}$ S1 $\xrightarrow{}$ S3 $\xrightarrow{}$ ES7  & 500 B & 4 ms \\ 
ES3 & V3 & ES6 & ES3 $\xrightarrow{}$ S2 $\xrightarrow{}$ S3 $\xrightarrow{}$ ES6  & 500 B & 4 ms \\ 
ES4 & V4 & ES6 & ES4 $\xrightarrow{}$ S2 $\xrightarrow{}$ S3 $\xrightarrow{}$ ES6  & 500 B & 4 ms \\ 
ES5 & V5 & ES6 & ES5 $\xrightarrow{}$ S3 $\xrightarrow{}$ ES6                   & 500 B & 4 ms \\ \hline
\end{tabular}}
\end{table}

The results of this use case are summarized in Table \ref{tab:check_results}. Each row of the table represents the worst delay of each \gls{vl}, and each \gls{es} transmission start time for achieving it, where $\Delta t$ is an insignificant delta of time used for establishing the packet order in the queues. As it can be observed in the three columns of the right, the simulation results match the results given in \cite{Xu2019}.

\begin{table}[h]
\centering
\caption{Testing use case delay results comparison.}
\label{tab:check_results}
\resizebox{\columnwidth}{!}{%
\begin{tabular}{c|ccccc|ccc|}
\cline{2-9}
\multicolumn{1}{l|}{} &
  \multicolumn{5}{c|}{\cellcolor[HTML]{9B9B9B} {\color[HTML]{FFFFFF} \textbf{Transmission start}}} &
  \multicolumn{3}{c|}{\cellcolor[HTML]{9B9B9B} {\color[HTML]{FFFFFF}\textbf{Evaluation Method}}} \\ \hline
  \rowcolor[HTML]{D0D0D0}
\multicolumn{1}{|c|}{\textbf{VL}} &
  \multicolumn{1}{c|}{\textbf{ES1}} &
  \multicolumn{1}{c|}{\textbf{ES2}} &
  \multicolumn{1}{c|}{\textbf{ES3}} &
  \multicolumn{1}{c|}{\textbf{ES4}} &
  \textbf{ES5} &
  \multicolumn{1}{c|}{\textbf{EPL}} &
  \multicolumn{1}{c|}{\textbf{BNCOG}} &
  \textbf{Simulation} \\ \hline
\multicolumn{1}{|c|}{\cellcolor[HTML]{EFEFEF}V1} &
  \multicolumn{1}{c|}{2$\Delta$ \text{t }  $\mu s$} &
  \multicolumn{1}{c|}{$\Delta$ \text{t }  $\mu s$} &
  \multicolumn{1}{c|}{0 $\mu s$} &
  \multicolumn{1}{c|}{0 $\mu s$} &
  96 $\mu s$ &
  \multicolumn{1}{c|}{272 $\mu s$} &
  \multicolumn{1}{c|}{272.8 $\mu s$} &
  272 $\mu s$ \\ 
\multicolumn{1}{|c|}{\cellcolor[HTML]{EFEFEF}V2} &
  \multicolumn{1}{c|}{0 $\mu s$} &
  \multicolumn{1}{c|}{$\Delta$ \text{t }  $\mu s$} &
  \multicolumn{1}{c|}{0 $\mu s$} &
  \multicolumn{1}{c|}{0 $\mu s$} &
  96 $\mu s$ &
  \multicolumn{1}{c|}{192 $\mu s$} &
  \multicolumn{1}{c|}{192 $\mu s$} &
  192 $\mu s$ \\ 
\multicolumn{1}{|c|}{\cellcolor[HTML]{EFEFEF}V3} &
  \multicolumn{1}{c|}{$\Delta$ \text{t }  $\mu s$} &
  \multicolumn{1}{c|}{0 µs} &
  \multicolumn{1}{c|}{2$\Delta$ \text{t }  $\mu s$} &
  \multicolumn{1}{c|}{$\Delta$ \text{t }  $\mu s$} &
  96 $\mu s$ &
  \multicolumn{1}{c|}{272 $\mu s$} &
  \multicolumn{1}{c|}{272.8 $\mu s$} &
  272 $\mu s$ \\ 
\multicolumn{1}{|c|}{\cellcolor[HTML]{EFEFEF}V4} &
  \multicolumn{1}{c|}{$\Delta$ \text{t }  $\mu s$} &
  \multicolumn{1}{c|}{0 $\mu s$} &
  \multicolumn{1}{c|}{$\Delta$ \text{t }  $\mu s$} &
  \multicolumn{1}{c|}{2$\Delta$ \text{t }  $\mu s$} &
  96 $\mu s$ &
  \multicolumn{1}{c|}{272 $\mu s$} &
  \multicolumn{1}{c|}{272.8 $\mu s$} &
  272 $\mu s$ \\ 
\multicolumn{1}{|c|}{\cellcolor[HTML]{EFEFEF}V5} &
  \multicolumn{1}{c|}{$\Delta$ \text{t }  $\mu s$} &
  \multicolumn{1}{c|}{0 $\mu s$} &
  \multicolumn{1}{c|}{0 $\mu s$} &
  \multicolumn{1}{c|}{0 $\mu s$} &
  96 + 2$\Delta$ \text{t } $\mu s$ &
  \multicolumn{1}{c|}{176 $\mu s$} &
  \multicolumn{1}{c|}{176.8 $\mu s$} &
  176 $\mu s$ \\ \hline
\end{tabular}%
}
\end{table}

\begin{figure}[h!]
\begin{adjustwidth}{-\extralength}{0cm}
    \begin{center}
        \subfloat[Switch 1.]{
        \includegraphics[width=0.6\textwidth]{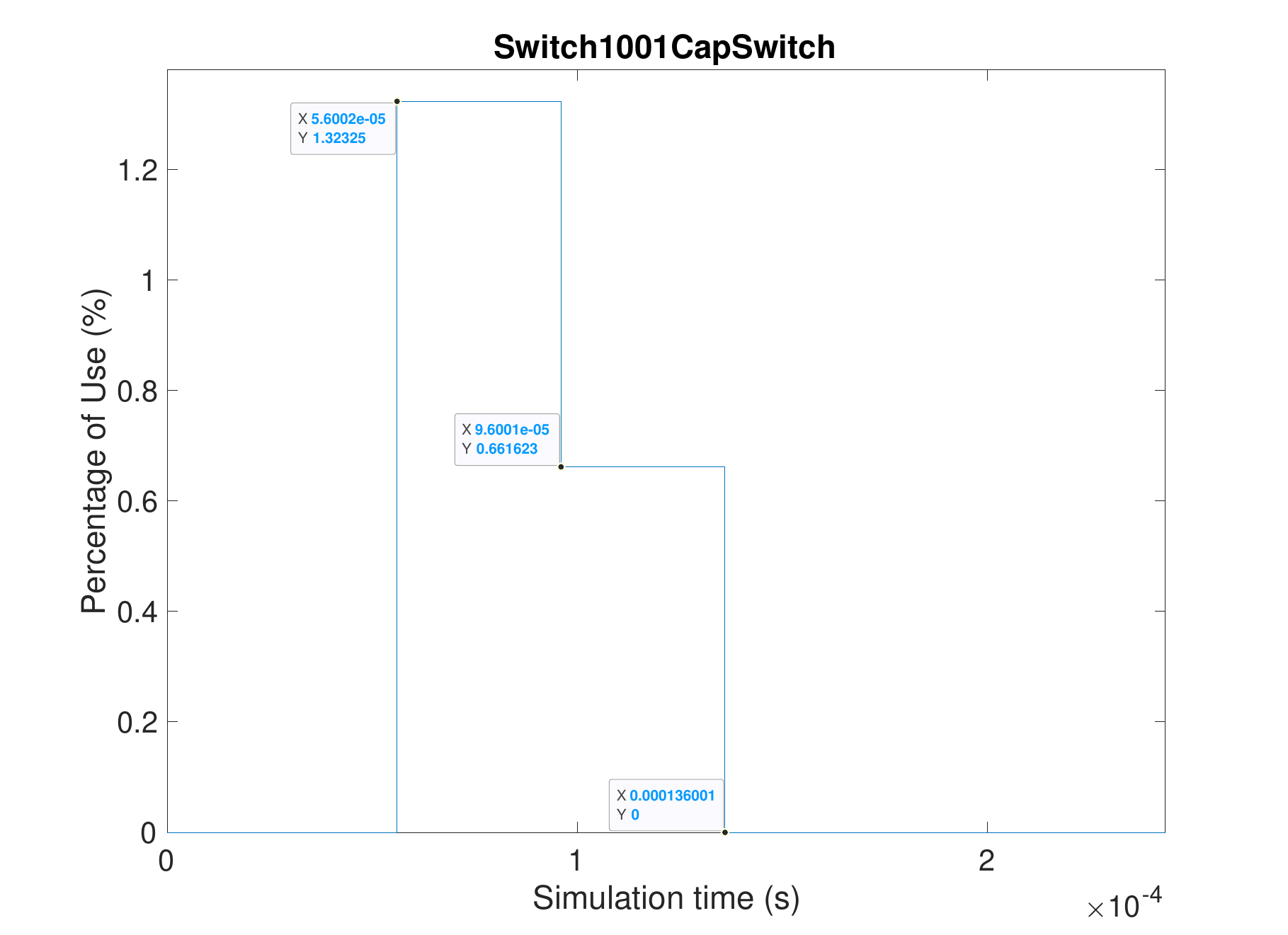}
        \label{fig:switch1}
        }
        \subfloat[Switch 2.]{
        \includegraphics[width=0.6\textwidth]{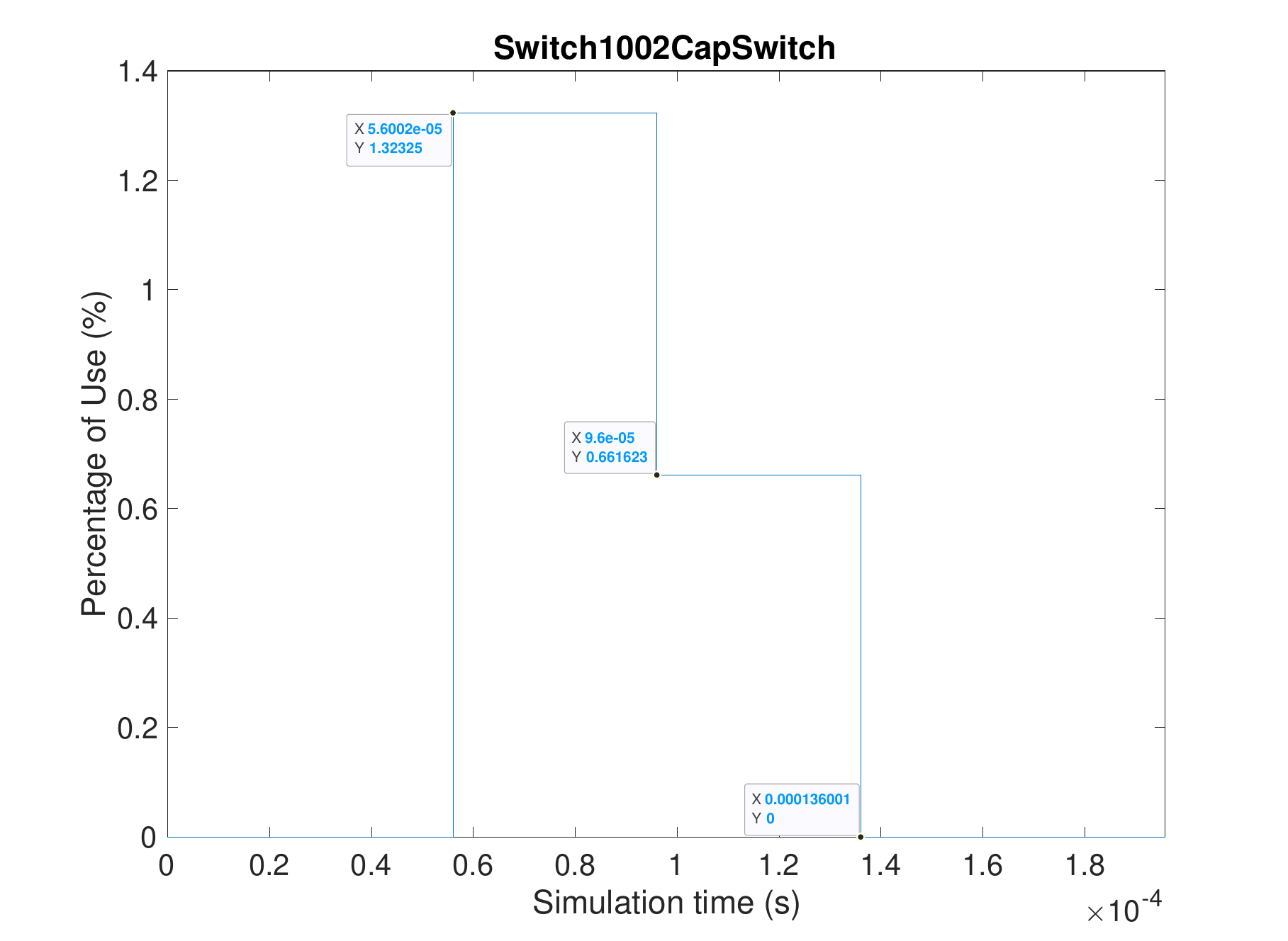}
        \label{fig:switch2}
        }\\
        \subfloat[Switch 3.]{
        \includegraphics[width=0.7\textwidth]{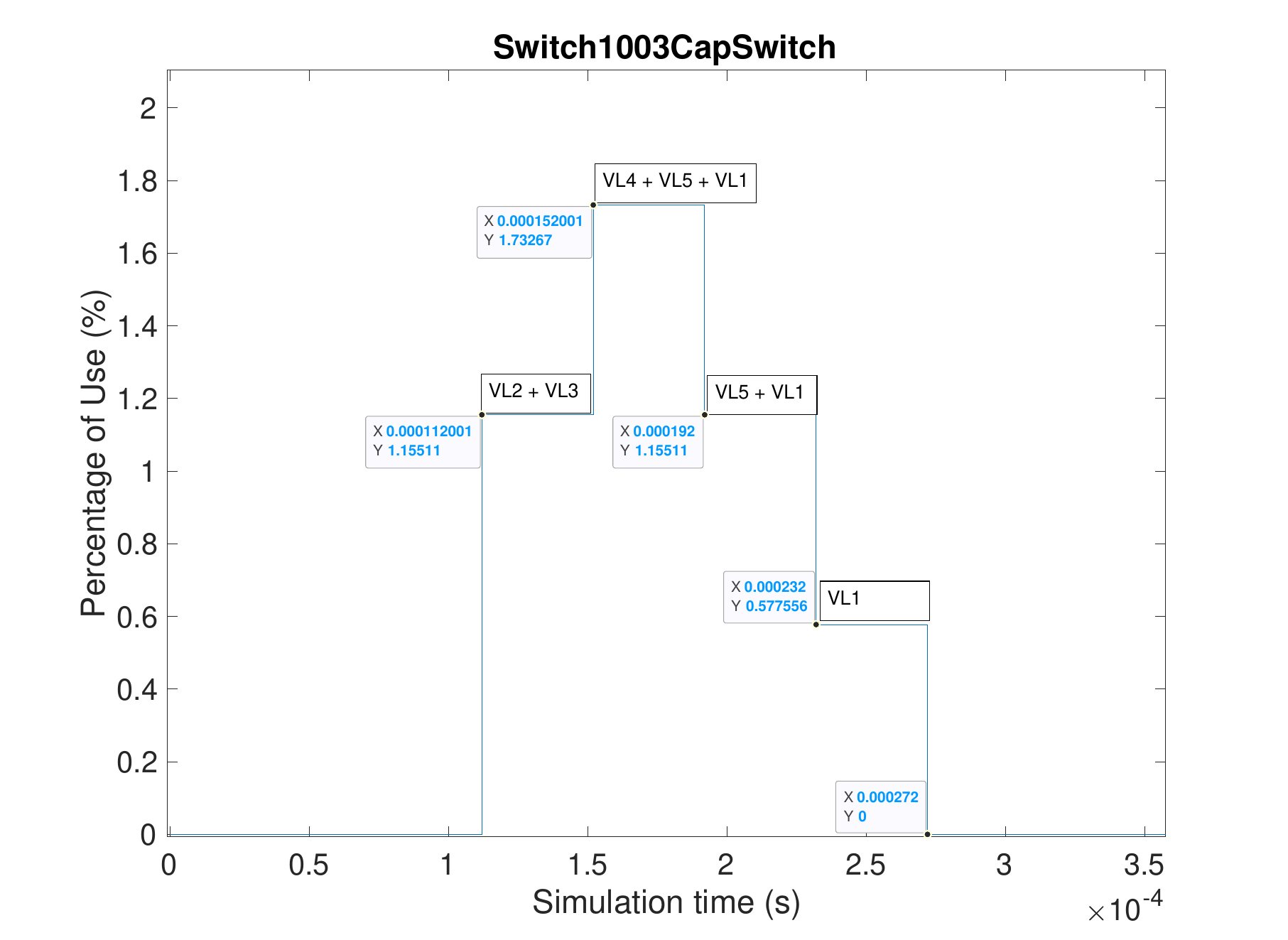}
        \label{fig:switch3}
        }
    \end{center}
\end{adjustwidth}   
\caption{Switch memory usage during packet collisions.}
\label{fig:Switches-capacity}
\end{figure}

Besides, the simulator gives the opportunity of study in depth the causes of the different delays by looking at the \gls{fom} of switch capacity. This can be extracted as the usage of the memory of the switch throughout the simulation. For example, the usage of memory of the three switches during the collision of packets is represented in Figure \ref{fig:Switches-capacity}. In particular, this simulation corresponds with the worst case for \gls{vl}1 presented in Table \ref{tab:check_results}. In both Switch 1 and Switch 2 (Figures \ref{fig:switch1} and \ref{fig:switch2}, respectively), it can be observed that two packets arrive at the same time and leave one after the other. Meanwhile, in Switch 3 (Figure \ref{fig:switch3}), more packets arrive at the same time: at time $t=112 \mu s$, the packets of \gls{vl}2 and \gls{vl}3 arrive to the switch and go to different queues; at time $t=152 \mu s$, both packets leave the switch and three packets from \gls{vl}4, \gls{vl}5 and \gls{vl}1 arrive to the same queue (being processed in that order); at time $t=192 \mu s$, the first packet in the queue leaves; at time $t=232 \mu s$, the second packet leaves; and, at time $t=272 \mu s$, the packet corresponding to \gls{vl}1 leaves and reaches its destination, as shown in Table \ref{tab:check_results}.

\subsection{Computational performance analysis}

\begin{figure}[h!]
    \centering
    \includegraphics[width = 0.85\textwidth]{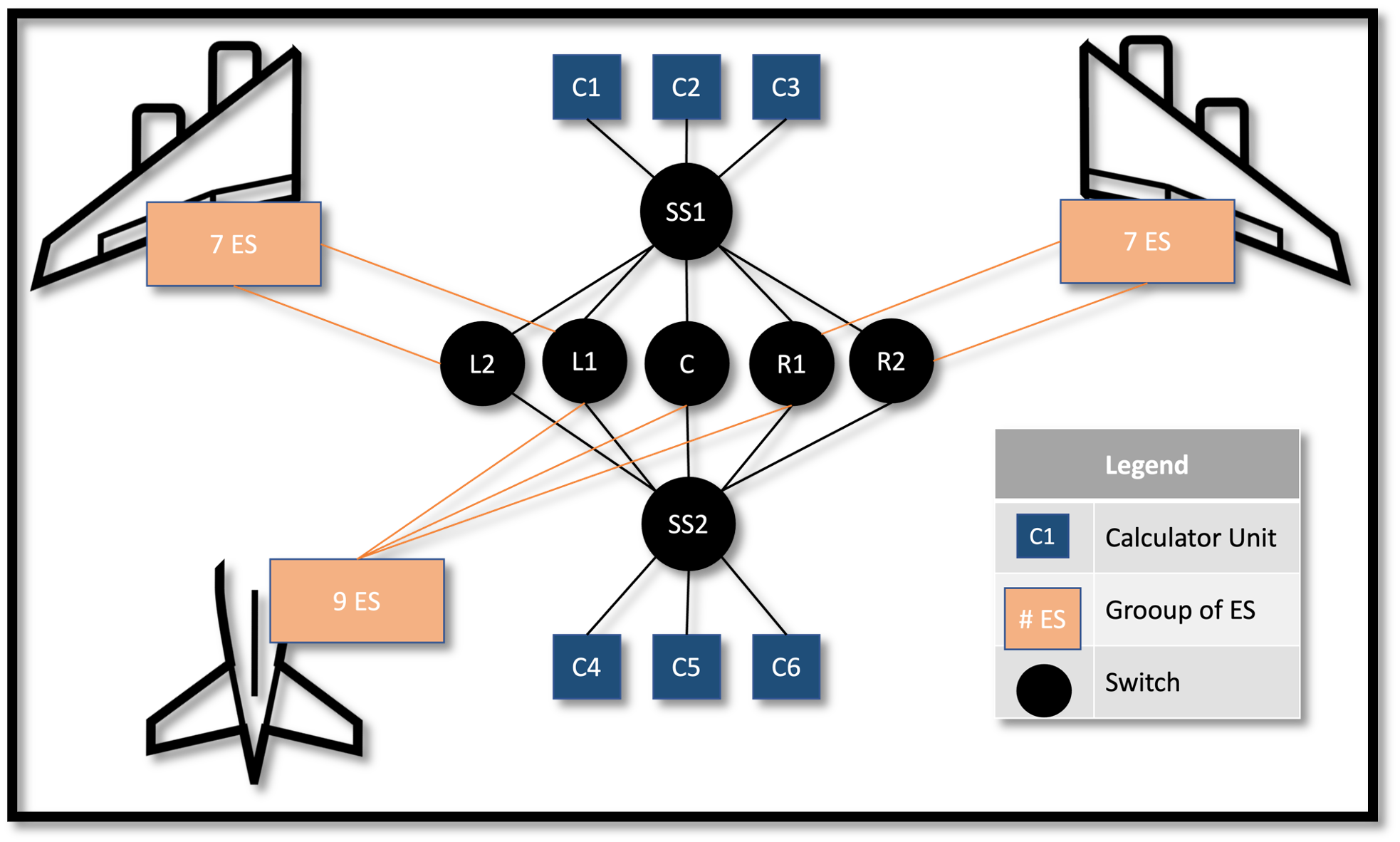}
    \caption{Airbus A350 architecture used for the performance analysis adapted from \cite{A350}.}
    \label{fig:A350-arch}
\end{figure}

In order to analyse the computational performance, the execution time of the simulations has been studied. For this, the network topology of a real airplane, such as the Airbus A350, has been simulated with different packet periodicity configurations. The Airbus A350 architecture, which has been adapted from \cite{A350}, is depicted in Figure \ref{fig:A350-arch}. This architecture is composed by 37 ES, from which 6 are \gls{cu}, and 7 switches.

\begin{figure}[h!]
    \centering
    \includegraphics[width=\textwidth]{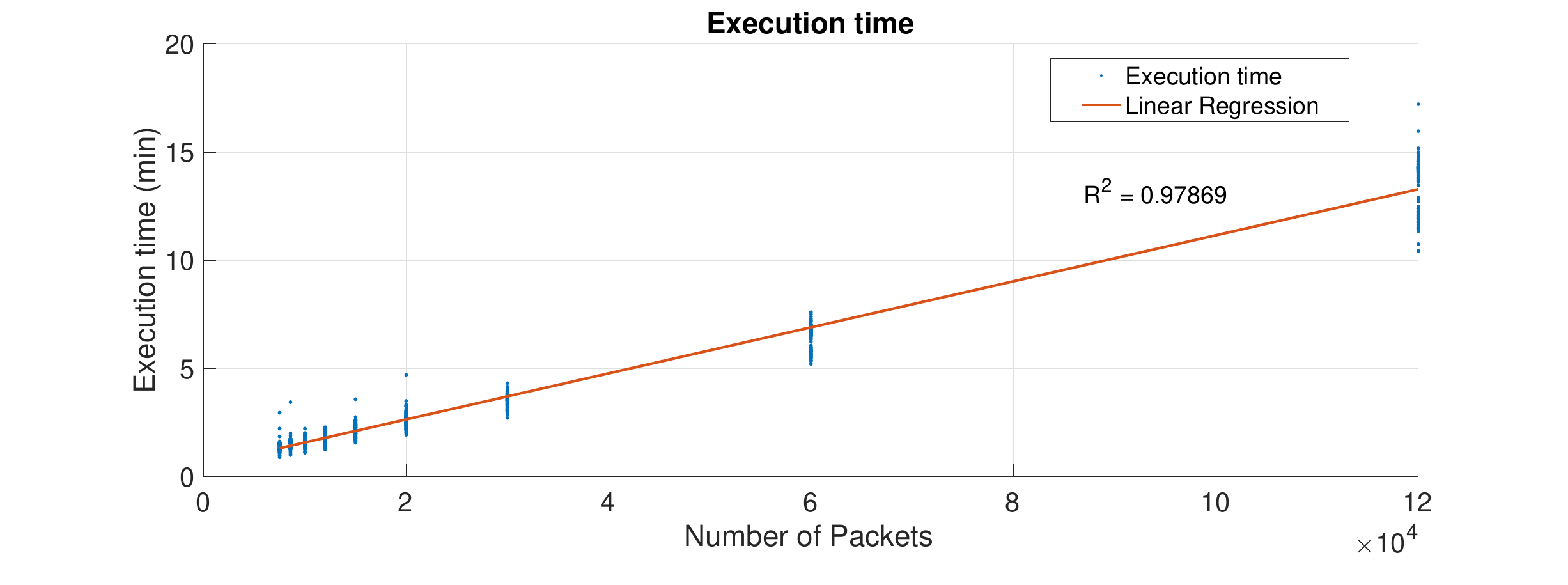}
\caption{Computational performance of the simulator: Execution time in minutes vs. Number of messages sent.}
\label{fig:time-number}
\end{figure}

Then, this topology has been simulated for 1 second with 60 \gls{vl} configured with a packet periodicity of 0.5 ms, 1 ms, 2 ms, 3 ms, 4 ms, 5 ms, 6 ms, 7 ms and 8 ms, meaning a total of $120000$, $60000$, $30000$, $20000$, $15000$, $12000$, $10000$, $8570$ and $7500$ messages sent, respectively. Each configuration has been simulated 100 times in order to obtain statistically significant results. The mean execution time of these configurations (running on a Mac with an Apple M1 chip and 16 GB of RAM) results in 13.5, 6.4, 3.5, 2.6, 2.2, 1.8, 1.6, 1.5 and 1.4 minutes, respectively, as shown in Figure \ref{fig:time-number}. These results show that the duration of the simulations has a clear linear dependence on the messages sent , resulting in the linear expression of Equation \ref{eq:1} with a correlation value of $R^2=0.97869$.

\begin{equation}
\label{eq:1}
    Ex\_time [min] = 0.000106 \cdot N\_Packets + 0.5
\end{equation}

In this way, it can be expected that the time required for each simulation will not grow much faster than the number of \gls{vl}. These execution times would allow real networks to be evaluated in a timely manner.

\begin{table}[!ht]
\caption{Delay statistics (in $\mu s$) of the A350 simulation with 0.5 ms of periodicity.}
    \centering
    \resizebox{\columnwidth}{!}{%
    \begin{tabular}{x{1cm}x{1cm}x{1.3cm}x{1cm}|x{1cm}x{1cm}x{1.3cm}x{1cm}|x{1cm}x{1cm}x{1.3cm}x{1cm}}
    \rowcolor[HTML]{9B9B9B}
    \hline
        {\color[HTML]{FFFFFF} \textbf{VL}} & {\color[HTML]{FFFFFF} \textbf{Mean}} & {\color[HTML]{FFFFFF} \textbf{Median}} & {\color[HTML]{FFFFFF}\textbf{Std}} & {\color[HTML]{FFFFFF}\textbf{VL}} & {\color[HTML]{FFFFFF}\textbf{Mean}} & {\color[HTML]{FFFFFF}\textbf{Median}} & {\color[HTML]{FFFFFF} \textbf{Std}} & {\color[HTML]{FFFFFF} \textbf{VL}} & {\color[HTML]{FFFFFF} \textbf{Mean}} & {\color[HTML]{FFFFFF} \textbf{Median}} & {\color[HTML]{FFFFFF} \textbf{Std}} \\ \hline 
        \cellcolor[HTML]{EFEFEF}1 & 67.18 & 65.52 & 4.44 & \cellcolor[HTML]{EFEFEF}21 & 109.42 & 115.3 & 31.77 & \cellcolor[HTML]{EFEFEF}41 & 99.71 & 98.1 & 21.41 \\ 
        \cellcolor[HTML]{EFEFEF}2 & 64.84 & 64.82 & 1.39 & \cellcolor[HTML]{EFEFEF}22 & 93.96 & 94.74 & 20.87 & \cellcolor[HTML]{EFEFEF}42 & 137.57 & 138.22 & 10.26 \\ 
        \cellcolor[HTML]{EFEFEF}3 & 86.61 & 84.97 & 17.4 & \cellcolor[HTML]{EFEFEF}23 & 81.17 & 76.19 & 15.52 & \cellcolor[HTML]{EFEFEF}43 & 91.55 & 94.79 & 17.11 \\ 
        \cellcolor[HTML]{EFEFEF}4 & 64.85 & 64.87 & 1.38 & \cellcolor[HTML]{EFEFEF}24 & 127.84 & 128.23 & 20.76 & \cellcolor[HTML]{EFEFEF}44 & 114.8 & 115.49 & 10.45 \\ 
        \cellcolor[HTML]{EFEFEF}5 & 70.26 & 73.72 & 6.1 & \cellcolor[HTML]{EFEFEF}25 & 163.94 & 161.97 & 5.02 & \cellcolor[HTML]{EFEFEF}45 & 74.88 & 74.53 & 6.69 \\ 
        \cellcolor[HTML]{EFEFEF}6 & 85.87 & 84.42 & 17.23 & \cellcolor[HTML]{EFEFEF}26 & 79.87 & 83.48 & 10.15 & \cellcolor[HTML]{EFEFEF}46 & 146.82 & 148.58 & 6.13 \\ 
        \cellcolor[HTML]{EFEFEF}7 & 100.96 & 97.86 & 26.72 & \cellcolor[HTML]{EFEFEF}27 & 70.33 & 66.44 & 6.97 & \cellcolor[HTML]{EFEFEF}47 & 125.13 & 119.35 & 16.12 \\ 
        \cellcolor[HTML]{EFEFEF}8 & 94.93 & 95.07 & 21 & \cellcolor[HTML]{EFEFEF}28 & 86 & 84.37 & 9.01 & \cellcolor[HTML]{EFEFEF}48 & 109.35 & 114.99 & 31.71 \\ 
        \cellcolor[HTML]{EFEFEF}9 & 69.31 & 67.18 & 5.17 & \cellcolor[HTML]{EFEFEF}29 & 69.65 & 66.25 & 6.05 & \cellcolor[HTML]{EFEFEF}49 & 102.25 & 105.66 & 23.66 \\ 
        \cellcolor[HTML]{EFEFEF}10 & 101.13 & 105.23 & 25.25 & \cellcolor[HTML]{EFEFEF}30 & 108.99 & 108.44 & 32.01 & \cellcolor[HTML]{EFEFEF}50 & 136.74 & 138.01 & 7.12 \\ 
        \cellcolor[HTML]{EFEFEF}11 & 100.64 & 105.27 & 24.69 & \cellcolor[HTML]{EFEFEF}31 & 91.56 & 85.98 & 19.67 & \cellcolor[HTML]{EFEFEF}51 & 124.96 & 119.36 & 15.77 \\ 
        \cellcolor[HTML]{EFEFEF}12 & 77.91 & 78.24 & 9.82 & \cellcolor[HTML]{EFEFEF}32 & 100.12 & 98.62 & 21.42 & \cellcolor[HTML]{EFEFEF}52 & 69.93 & 66.86 & 6.08 \\ 
        \cellcolor[HTML]{EFEFEF}13 & 103 & 97.31 & 27.47 & \cellcolor[HTML]{EFEFEF}33 & 115.84 & 115.89 & 20.84 & \cellcolor[HTML]{EFEFEF}53 & 84.21 & 84.7 & 15.06 \\ 
        \cellcolor[HTML]{EFEFEF}14 & 94.74 & 96.23 & 14.12 & \cellcolor[HTML]{EFEFEF}34 & 78.22 & 83.3 & 9.59 & \cellcolor[HTML]{EFEFEF}54 & 99.89 & 97.91 & 10.37 \\ 
        \cellcolor[HTML]{EFEFEF}15 & 100.26 & 105 & 25.22 & \cellcolor[HTML]{EFEFEF}35 & 116.18 & 116.3 & 21.24 & \cellcolor[HTML]{EFEFEF}55 & 114.76 & 115.55 & 21.09 \\ 
        \cellcolor[HTML]{EFEFEF}16 & 113.15 & 115.19 & 8.21 & \cellcolor[HTML]{EFEFEF}36 & 109.96 & 114.91 & 31.68 & \cellcolor[HTML]{EFEFEF}56 & 159.96 & 160.07 & 8.61 \\ 
        \cellcolor[HTML]{EFEFEF}17 & 150.42 & 151.24 & 11.63 & \cellcolor[HTML]{EFEFEF}37 & 115.58 & 117.89 & 7.11 & \cellcolor[HTML]{EFEFEF}57 & 142.03 & 140.26 & 7 \\ 
        \cellcolor[HTML]{EFEFEF}18 & 123.98 & 125.67 & 15.06 & \cellcolor[HTML]{EFEFEF}38 & 168.25 & 170 & 5.05 & \cellcolor[HTML]{EFEFEF}58 & 135.14 & 138.07 & 7.06 \\ 
        \cellcolor[HTML]{EFEFEF}19 & 86.69 & 84.96 & 17.28 & \cellcolor[HTML]{EFEFEF}39 & 99.98 & 96.41 & 26.36 & \cellcolor[HTML]{EFEFEF}59 & 142.2 & 139.63 & 5.79 \\ 
        \cellcolor[HTML]{EFEFEF}20 & 101.2 & 105.55 & 24.6 & \cellcolor[HTML]{EFEFEF}40 & 69.99 & 73.08 & 6.09 & \cellcolor[HTML]{EFEFEF}60 & 115.45 & 117.68 & 7.06 \\ \hline
    \end{tabular}
    }
    \label{tab:statistics_delay}
\end{table}


Furthermore, the simulation-derived packet traces can serve as valuable data for generating time series metrics. This feedback is crucial in the design process and provides insight into the performance of the network during regular operation. It allows the evaluation of the efficiency of the network and the establishment of metrics beyond the worst-case delays that are typically used for certification purposes.

In this way, in the table \ref{tab:statistics_delay} the statistics of the simulation for the periodicity of 0.5 $\mu s$. In the configuration of the simulations, a small difference of 200 Bytes between the minimum and maximum packet length has been set so that there is a variance in the traces of the \gls{vl}s. The 60 \gls{vl}s can be divided in two groups: the ones that go from a \gls{cu} to a \gls{es} (\gls{vl}s 2,4,5,29,40 and 52) and the ones that go from a \gls{es} to a \gls{cu} (the rest of \gls{vl}s). As it can be observed, the first group of \gls{vl}s have a similar low latency (around 65 $\mu s$) due to that there are only five and the probabilities of conflicts in the queues are low. Meanwhile, the rest of \gls{vl}s have delays from 65 $\mu s$ to 170 $\mu s$ due to the high probability of matching queues. Also, some \gls{vl}s depart from the same \gls{es}, making their packets to wait until the previous one is sent.

\section{Conclusions and outlook}
\label{chp:conclusions}

In this work, a short analysis of the main protocols and standards used to communicate the different elements in an avionics network, such as Ethernet and Ethernet-based \gls{afdx} and \gls{tsn}, has been presented. Also, it has bee observed that there is an effort to substitute \gls{afdx} with more low-cost Ethernet-based devices in the market. Moreover, it has been determined that the \gls{sil} step of \gls{mbse} design process is of great importance in the aerospace sector and, in concordance, a Matlab/Simulink-based link-level simulator for Ethernet and \gls{afdx} has been developed as a relevant tool for the early stages of the design process. This simulator can be easily integrated into validation frameworks and platforms. Besides, it has been successfully tested by replicating the results of a known use case, showing the possibilities of analysis that the simulator’s \gls{fom}s procure as well. Then, the computational performance analysis with a real use case of Airbus A350 network topology has shown that the time needed for each simulation has a linear dependence with the number of messages sent, allowing evaluating real networks in a timely manner. Also, the fact that the simulator has been designed as an event-driven simulator makes it more efficient than other simulators with fixed-step solvers, and the versatility of the results facilitates informed decision making and refinement of avionics networks.




Further work would include the study of the \gls{tsn} standards for their implementation in the simulator, as \gls{tsn} is anticipated to become the standard for future generations of aircraft. Also, validation frameworks and platforms would be developed in order to automate the avionics design process.



\vspace{6pt} 

\authorcontributions{Conceptualization, P.V., J.V. and S.F.; methodology, P.V. and J.V.; software, P.V., J.V. and J.P.; validation, P.V., J.V. and V.E.; formal analysis, J.V.; writing---original draft preparation, P.V.; supervision, S.F.; project administration, S.F. and R.B.; funding acquisition, S.F., V.E., R.O. and R.B. All authors have read and agreed to the published version of the manuscript.}

\acknowledgments{This work has been partially funded by: 
the Junta de Andalucía and the ERDF(European Regional Development Fund) Operational Programme in the framework of the CAPTOR project: “advanCed Avionics communications validation and verification PlaTfORm” (Ref. PYC20 RE 077 UMA); 
AERTEC Solutions (reference 8.06/5.59.5543, 8.06/5.59.5715, 806/59.5974 and 806/59.5715) in the framework of project 2020 AS-DISCO: “Audio Suite for Disruptive Cockpit Demonstrator” (this project has received funding from the Clean Sky 2 Joint Undertaking under the European Union's Horizon 2020 research and innovation programme under Grant Agreement n°: 865416’); 
Ministry of Economic Affairs and Digital Transformation and the European Union - NextGenerationEU, in the framework of the Recovery, Transformation and Resilience Plan and the Recovery and Resilience Mechanism under the MAORI project; 
the Ministry of Science and Innovation (grant FPU21/04472); and the University of Malaga through the “II Plan Propio de Investigación, Transferencia y Divulgación Científica”.
The authors are grateful to Aertec Solutions for their support and collaboration in this project.}


\conflictsofinterest{The authors declare no conflict of interest.}




\printglossary[type=\acronymtype]




\begin{adjustwidth}{-\extralength}{0cm}

\reftitle{References}
\bibliography{EASN}

\PublishersNote{}
\end{adjustwidth}
\end{document}